\documentclass[draftcls,onecolumn]{IEEEtran}

\usepackage{epsfig}

\begin{document}
\title{Simulation of Graphene Nanoribbon Field Effect Transistors}

\author{\normalsize Gianluca Fiori, Giuseppe Iannaccone\\
 Dipartimento di Ingegneria dell'Informazione~: Elettronica, Informatica, Telecomunicazioni,\\
Universit\`a di Pisa, Via Caruso 16, 56126 Pisa, Italy.\\
email : g.fiori@iet.unipi.it; Tel. +39 050 2217638; Fax : + 39 050 2217522\\}
\maketitle
\newpage
\vspace{5cm}
\begin{abstract}
We present an atomistic three-dimensional simulation of graphene nanoribbon field 
effect transistors (GNR-FETs), based on the self-consistent solution of the 3D Poisson 
and Schr\"odinger equation with open boundary conditions within the non-equilibrium 
Green's Function formalism and a tight-binding Hamiltonian. 
With respect to carbon nanotube FETs, GNR-FETs exhibit comparable performance, 
reduced sensitivity on the variability of channel chirality, and similar 
leakage problems due to band-to-band tunneling.
Acceptable transistor performance requires prohibitive effective nanoribbon width of 1-2 nm and 
atomistic precision, that could in principle be obtained with periodic etch patterns or stress patterns.
\end{abstract}

{\bf Keyworks:} graphene, nanoribbon, NEGF, three-dimensional Poisson, Atomistic tight-binding Hamiltonian.
\newpage

\section{Introduction}

In the last decade, Carbon NanoTubes (CNT) have attracted extraordinary interest 
for their extremely interesting physical and electrical properties~\cite{Javey}, and their potential
as an alternative to silicon as channel material for transistors beyond CMOS technology~\cite{ITRS}.
 Recent experiments by Novoselov et al.~\cite{Novoselov} demonstrated the 
possibility of fabricating stable single atomic layer graphene sheets, 
with remarkable electrical properties, that have brought new excitation 
to the field of carbon electronics.

Two-dimensional graphene is a zero gap material, which makes it not suitable for transistor applications.

Energy gap can however be induced by means of lateral confinement~\cite{Han}, 
realized for example by lithography definition of narrow graphene stripes, the so-called
graphene nanoribbons.

Experiments on graphene-based devices~\cite{Kurz} and Graphene NanoRibbon FETs~\cite{Avouris} 
(GNR-FETs) have appeared only very recently, and demonstrate limited capability to
modulate the conductance of a graphene channel at room temperature.
The main problem is the need to fabricate extremely narrow nanowires (of the order of 1 nm) with atomic precision 
to obtain an energy gap adequate for room temperature operation.

Since at the moment the fabrication technology is at its very first steps, computer simulations
can be very useful to provide physical insights of GNR-FETs and to estimate the attainable performance. 
Recent theoretical works have shown that graphene nanoribbons have an energy gap which has an oscillating
behavior as a function of width, with average
roughly proportional to the inverse width, and that edge states play a very important role 
in inhibiting the existence of fully metallic nanoribbons~\cite{Cohen}. Such behavior cannot be reproduced if one does
not consider edge effects~\cite{ventidue}.

Also from the simulation point of view, research on GNR-FETs is at an embryonic stage:
the only works available in the literature~\cite{Jing,Albert}
are based on a semiclassical analytical top-of-the-barrier model.
For short-channel transistors only a three-dimensional simulation is suitable for an accurate evaluation
of the electrostatics and of intraband and interband tunneling.

To this purpose, we have developed a code for the simulation of GNR-FETs,
based on the Non-Equilibrium Green's Function formalism (NEGF), with a tight-binding Hamiltonian
built from $p_z$ orbital basis set
in the real space, which has been included in our in-house three-dimensional device simulator NANOTCAD ViDES
~\cite{ViDES}. We will show that GNR-FETs have performance comparable with CNT-FETs, 
and that can be greatly affected by the channel width and edge roughness.

\section{Physical model and Results}

Our approach is based on the self-consistent solution of the three-dimensional
Poisson and Schr\"odinger equations with open boundary conditions~\cite{noi},
which is able to take into account fully ballistic transport, in order to
outline the higher limits of device performance, as well as elastic scattering due
to line edge roughness.
The Hamiltonian is taken from~\cite{Cohen}, in which edge states
at the nanoribbon lateral ends have been considered.
In this work we refer to ($N$,0) armchair graphene nanoribbons, which consist of an
unrolled ($n$,0) zig-zag nanotube with $N=2n$.

The considered double-gate GNR-FETs have the structure depicted in the inset of Fig.~\ref{fig2}.
The gates are metallic, the oxide thickness $t_{ox}$ is equal to 1~nm,
the channel is 15~nm long, and $W$ is the channel width.
 The source and drain extensions are 10~nm long, and are doped with a molar fraction of fully ionized
donors $f=5\times10^{-3}$. The spacing between parallel GNRs is 4~nm.

In Fig.~\ref{fig2} the transfer characteristics of a (12,0) GNR-FET ($W$=1.37~nm) 
for drain-to-source voltage $V_{DS}$ of 0.1 and 0.5~V are shown, and 
compared to those of a (16,0) CNT-FET with the same geometry
 (same $t_{ox}$, $L$ and device spacing),  whose energy gap ($E_{gap}$) is close to that of the GNR-FET and
equal to 0.6~eV.

 Good control of the channel by the gate potential is shown at $V_{DS}$=~0.1~V,
since the sub-threshold swing ($S$)  for the GNR-FET and the CNT-FET are 64 and 68 mV/dec, respectively.
For $V_{DS}$=~0.5~V we observe a pronounced degradation of  
$S$, with $S$=191~mV/dec for the GNR-FET and almost 250~mV/dec for the CNT-FET.
This has to be imputed to Hole-Induced Barrier Lowering (HIBL)~\cite{noi}~:
 in the sub-threshold regime, when sufficiently high $V_{DS}$ is applied, 
confined states in the valence band of the channel align with the occupied states in the drain, leading 
to band-to-band injection of holes
in the channel.

If only elastic band-to-band tunneling can occur (as assumed in our simulation)
the excess of holes in the channel lowers the channel potential,
increasing the off current and $S$, as shown in Fig.~\ref{fig2}: the lower the energy gap and the
higher the $V_{DS}$, the higher the HIBL effect. 
HIBL is more pronounced in CNT-FETs than in GNR-FETs, because the conduction band of CNTs is double degenerate
and therefore CNTs have twice the density of states of GNRs with the same gap.

If, on the other hand, inelastic band-to-band tunneling or Schockley-Read-Hall mechanisms are relevant,
holes can recombine with electrons at the source and, instead of HIBL, we observe
a leakage current from source to drain due to gate-induced drain leakage (GIDL)~\cite{Chen}.

In strong inversion, the transconductance $g_m$ at $V_{DS}$=0.1~V is 3600 and 6100~$\mu$S/$\mu$m
for the GNR-FET and the CNT-FET, respectively, whereas at $V_{DS}$=0.5~V, we obtain
$g_m$=4800~$\mu$S/$\mu$m for the GNR and a
$g_m$=8760~$\mu$S/$\mu$m for the CNT. The advantage of CNT-FETs is due to the 
double degeneracy of the conduction band in carbon nanotubes.

It is known that a variability of the chirality of fabricated CNTs yields metallic nanotubes useless for
transistor applications. For GNRs this problem is mitigated, since all GNRs are semiconducting.
In order to investigate quantitatively the effect of a finite fabrication tolerance on the width of GNRs, we
have computed the transfer characteristics of GNR-FETs with different chiralities~: (12,0), (14,0) and (16,0).

As can be seen in Fig.~\ref{trans}a, the three devices behave as transistors, 
but show very different behavior, even if they differ 
by only one carbon atom along the channel width. The problem is that the gap is still largely dependent on 
the chirality~:~ the (16,0) GNR ($W$=1.87~nm) has the larger gap ($E_{gap}=0.71$~eV), while 
the (14,0) ($W$=1.62~nm) has the smallest gap ($E_{gap}=0.13$~eV).
 As a consequence, the (16,0) device show the best gate control over the
channel potential, while the (14,0) the worst~: the energy gap is so small that
elastic band-to-band tunneling occurs at the source and current is dominated by GIDL.

Such problem is reduced if rough edges are considered. We have considered the impact of line edge roughness in 
a (16,0) GNR-FET device, by randomly decoupling carbon atoms on the lateral boundaries of the GNR.
The transfer characteristic for one example is shown in Fig.~\ref{trans} (dashed line). 
Since the channel consists of several
hundreds of rings, the rough GNR behaves as a GNR with an intermediate effective gap. More statistical simulations
would be needed to assess the dispersion of the electrical characteristics, but the typical GNRs is probably
long enough to provide sufficient averaging to suppress inter device dispersion. 
Rough edge scattering strongly affects the on-current  and the transconductance suppressing it
by about 30\% with respect to fully ballistic transistors. 
Additional suppression in realistic GNR-FETs can be due to defects, ionized impurities and 
phonon scattering.

From the above simulations, it is clear that lateral confinement way
beyond state-of-the-art etching capabilities would be needed to obtain adequate
$E_{gap}$. 
We also found that electrostatic periodic potential modulation with a peak-to-peak value of 
few Volts is not sufficient to induce the required gap of few hundreds mV.

In order to evaluate whether a periodic strain pattern can allow to engineer the GNR gap,
we have computed the energy gap in a (24,0) GNR ($W$=5.86~nm), multiplying the 
overlap integral of the element of the Hamiltonian in 
correspondence of the couple of atoms in the middle of the GNR by a "strain factor" $\sigma$~: 
$\sigma$ is larger than one for compressive strain, 
and smaller than one for tensile strain. 
As can be seen in Fig.~\ref{ilgap}, compressive strain seems to be able to increase
the energy gap of the nanowire by a significant amount. Of course, we can
only suggest to experimentalists to evaluate the option. 

\section{Conclusions}

In this work, a simulation study of GNR-FETs has been performed by means of the self-consistent
solution of the 3D Poisson and Schr\"odinger equation with open boundary conditions, within the 
NEGF formalism.
 Edge states have been considered at the lateral ends of the nanoribbon using
the model proposed by~\cite{Cohen}.
GNR-FETs exhibit performance similar to CNT-FETs, also showing significant band-to-band
tunneling when small gap devices are considered and 
large $V_{DS}$ is applied. 
GNR-FETs are more robust than CNT-FETs with respect to variability of the channel chirality,
and edge roughness seems to play an useful averaging effect. Finally we 
suggest that periodic strain could in principle represent an alternative to etching 
for inducing an energy gap in graphene.

\section*{Acknowledgment}

Authors thank Prof. Massimo Macucci for suggestions and fruitful discussions.
Support from the European Science Foundation  EUROCORES Programme Fundamentals of Nanoelectronics, through
funding from the CNR (awarded to IEEIIT-PISA) and the EC Sixth Framework Programme, under project 
Dewint (Contract N. ERAS-CT-2003-980409) is gratefully acknowledged.

\newpage

\newpage

%************************** CAPTIONS *****************************************************

\begin{figure}
\caption{Transfer characteristics of double-gate CNT and GNR-FETs, with doped source and drain 
reservoirs, with channel length equal to 15~nm, oxide thickness $t_{ox}$ equal to 1~nm and channel width
$W$=1.37~nm.} The lateral space is equal to 2~nm. In the inset, a sketch of the GNR-FET is shown.
\label{fig2}	
\caption{Transfer characteristics in the logarithmic a) and linear b) scale 
of GNR-FETs with different chiralities (12,0), (14,0) and (16,0) (channel width $W$
equal to 1.37~nm, 1.62~nm and 1.87~nm, respectively), for
$V_{DS}$=0.1~V. The transfer characteristic for the (16,0) GNR-FET, when roughness at the lateral edge 
of the GNR is considered is also shown (dashed line). In the inset, a sketch of the graphene nanoribbon
is shown, where randomly decoupled atoms have been highlighted (thick lines).
}
\label{trans}	
\caption{Energy gap of a (24,0) GNR, when tensile and compressive strain is considered in correspondence of the
middle of the nanoribbon, as a function of the strain factor by which the Hamiltonians elements
of the strained carbon atoms are multiplied.}
\label{ilgap}	
\end{figure}

%************************** FIGURES AND TABLES **********************

\begin{figure}
\begin{center}
\vspace{8cm}

\epsfig{file=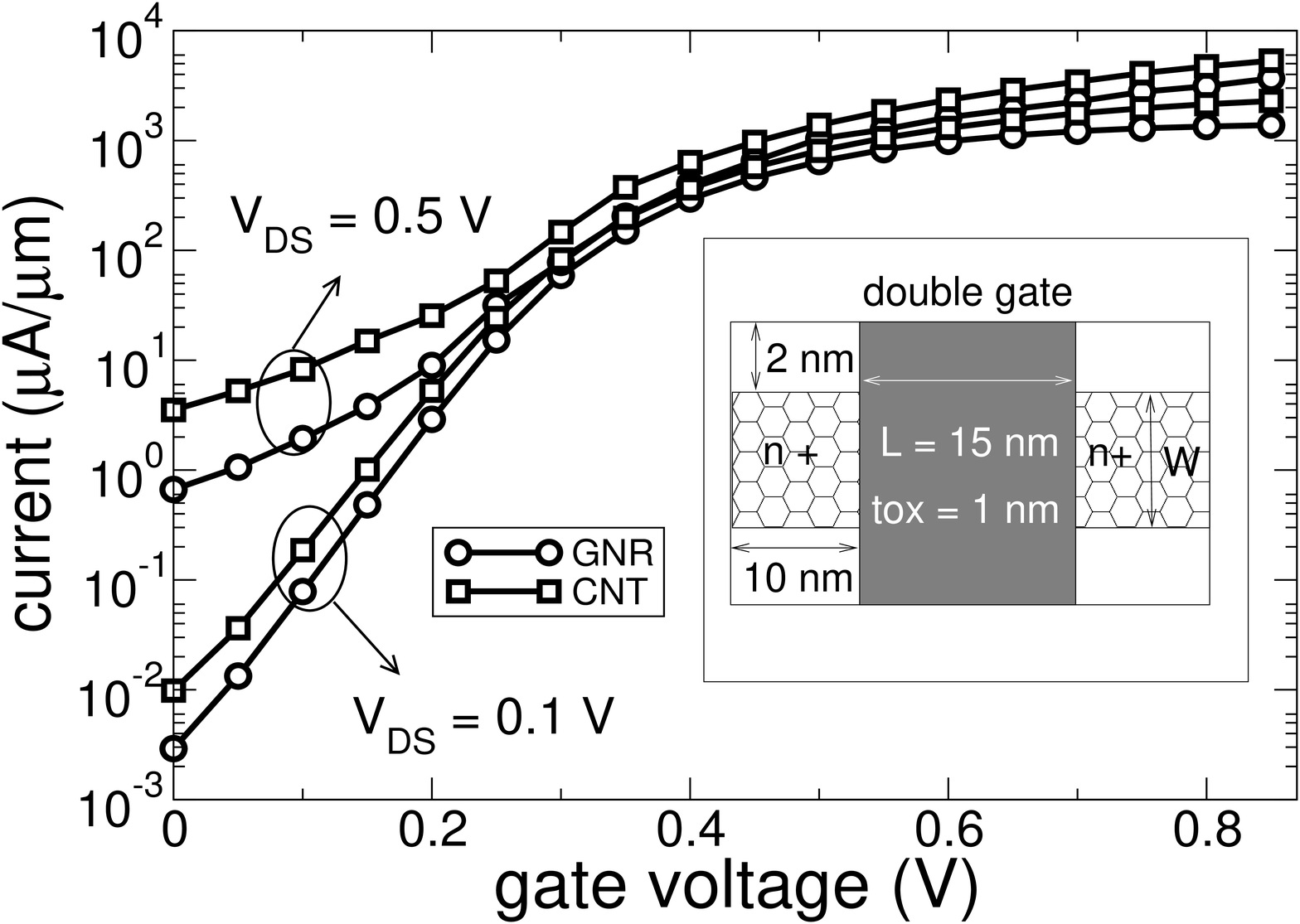,width=12cm} 
\huge
\vspace{1cm}\\
FIG. 1\\
Gianluca Fiori, Giuseppe Iannaccone \\
Electron Device Letters\\
\normalsize	
\end{center}
\end{figure}

\begin{figure}
\begin{center}
\epsfig{file=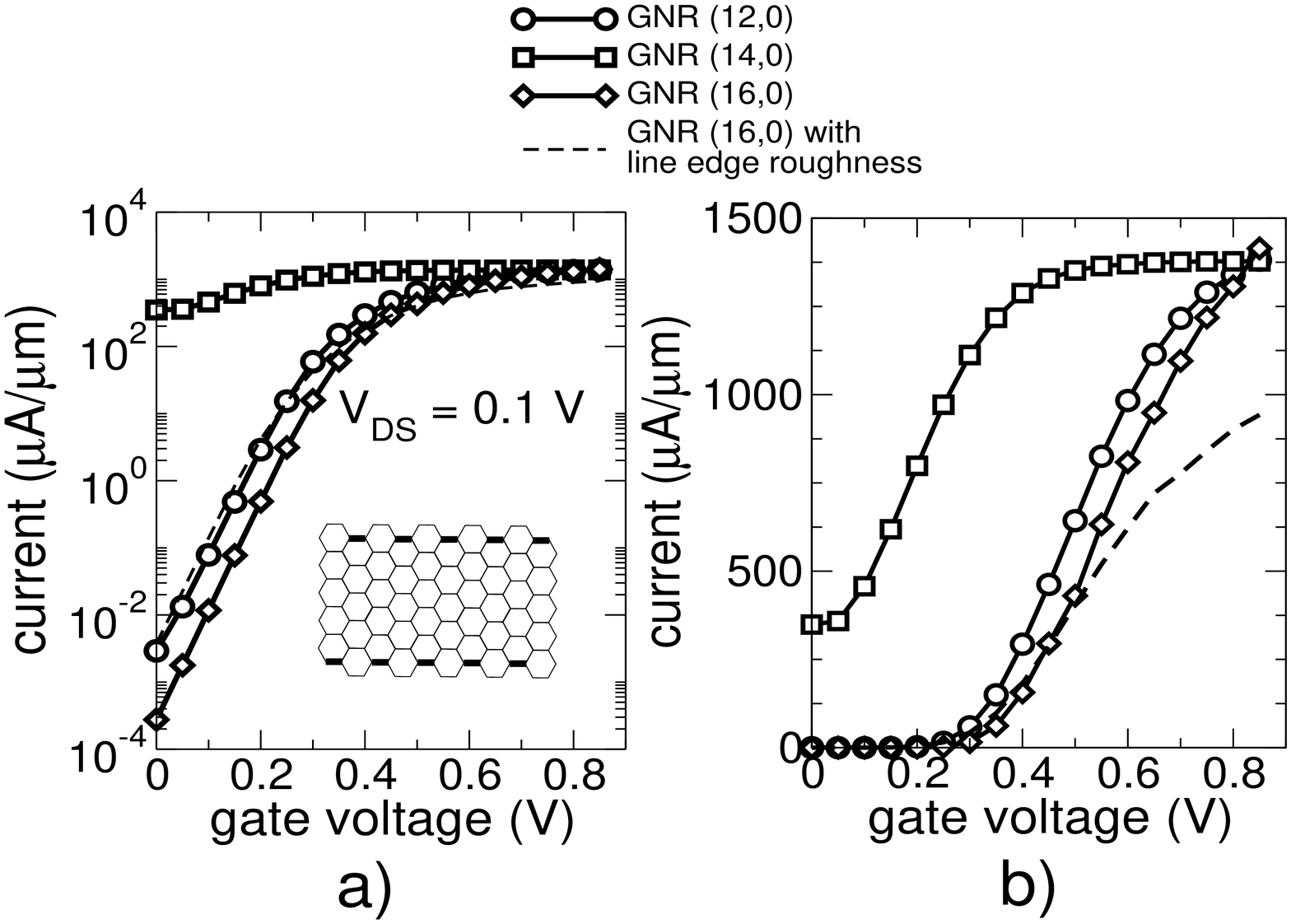,width=14cm} 
\huge
\vspace{1cm}\\
FIG. 2\\
Gianluca Fiori, Giuseppe Iannaccone \\
Electron Device Letters\\
\normalsize	
\end{center}
\end{figure}

\begin{figure}
\begin{center}
\epsfig{file=./Fig3.eps,width=12cm} 
\huge
\vspace{1cm}\\
FIG. 3\\
Gianluca Fiori, Giuseppe Iannaccone \\
Electron Device Letters\\
\normalsize	
\end{center}
\end{figure}

\end{document}